\RequirePackage{fix-cm}
\documentclass[pdftex,smallextended]{svjour3}       
\smartqed  
\usepackage{graphicx}
\usepackage{rotating}
\usepackage{natbib}
\usepackage[breaklinks=true]{hyperref}
\usepackage[all]{hypcap}
\renewcommand{\url}[1]{\href{#1}{#1}}
\newcommand{\aw}{\textsf{Astro-WISE}}
\newcommand{\awinfsys}{\textsf{Astro-WISE Information System}}
\newcommand{\qwise}{\textsf{Quality-WISE}}
\newcommand{\awe}{\texttt{AWE}}
\newcommand{\awenv}{\aw\ \textsf{Environment}}
\newcommand{\aweprompt}{\texttt{awe}-prompt}
\newcommand{\readn}{\texttt{\href{http://doc.astro-wise.org/astro.main.ReadNoise.html}{Read\-Noise}}}
\newcommand{\gainl}{\texttt{\href{http://doc.astro-wise.org/astro.main.GainLinearity.html}{Gain\-Linearity}}}
\newcommand{\rawbi}{\texttt{\href{http://doc.astro-wise.org/astro.main.RawBiasFrame.html}{Raw\-Bias\-Frame}}}
\newcommand{\rawdo}{\texttt{\href{http://doc.astro-wise.org/astro.main.RawDomeFlatFrame.html}{Raw\-Dome\-Flat\-Frame}}}
\newcommand{\rawtw}{\texttt{\href{http://doc.astro-wise.org/astro.main.RawTwilightFlatFrame.html}{Raw\-Twilight\-Flat\-Frame}}}
\newcommand{\biasf}{\texttt{\href{http://doc.astro-wise.org/astro.main.BiasFrame.html}{Bias\-Frame}}}
\newcommand{\twili}{\texttt{\href{http://doc.astro-wise.org/astro.main.TwilightFFlatFrame.html}{Twilight\-Flat\-Frame}}}
\newcommand{\domef}{\texttt{\href{http://doc.astro-wise.org/astro.main.DomeFlatFlatFrame.html}{Dome\-Flat\-Frame}}}
\newcommand{\maste}{\texttt{\href{http://doc.astro-wise.org/astro.main.MasterFlatFrame.html}{Master\-Flat\-Frame}}}
\newcommand{\coldp}{\texttt{\href{http://doc.astro-wise.org/astro.main.ColdPixelMap.html}{Cold\-Pixel\-Map}}}
\newcommand{\hotpi}{\texttt{\href{http://doc.astro-wise.org/astro.main.HotPixelMap.html}{Hot\-Pixel\-Map}}}
\newcommand{\reduc}{\texttt{\href{http://doc.astro-wise.org/astro.main.ReducedScienceFrame.html}{Reduced\-Science\-Frame}}}
\newcommand{\astro}{\texttt{\href{http://doc.astro-wise.org/astro.main.AstrometricParameters.html}{Astro\-metric\-Parameters}}}
\newcommand{\photo}{\texttt{\href{http://doc.astro-wise.org/astro.main.PhotometricParameters.html}{Photo\-metric\-Parameters}}}
\newcommand{\regri}{\texttt{\href{http://doc.astro-wise.org/astro.main.RegriddedFrame.html}{Regridded\-Frame}}}
\newcommand{\coadd}{\texttt{\href{http://doc.astro-wise.org/astro.main.CoaddedRegriddedFrame.html}{Coadded\-Regridded\-Frame}}}
\newcommand{\weigh}{\texttt{\href{http://doc.astro-wise.org/astro.main.WeightFrame.html}{Weight\-Frame}}}
\newcommand{\assoc}{\texttt{\href{http://doc.astro-wise.org/astro.main.AssociateList.html}{Associate\-List}}}
\newcommand{\proce}{\texttt{\href{http://doc.astro-wise.org/astro.main.ProcessTarget.html}{Process\-Target}}}

\journalname{Experimental Astronomy}
\begin{document}

\title{The \aw\ approach to quality control for astronomical data}

\subtitle{}

\titlerunning{\aw\ quality control}        

\author{
        McFarland, J.P.       \and
        Helmich, E.M.         \and
        Valentijn, E.A.
}

\authorrunning{McFarland, et al.} 

\institute{OmegaCEN, Kapteyn Astronomical Institute, Groningen University
           Postbus 800, 9700 AV, Groningen, The Netherlands\\
           \email{mcfarland@astro.rug.nl}
          }

\date{Received: date / Accepted: date}

\maketitle

\begin{abstract}
We present a novel approach to quality control during the processing of
astronomical data.  Quality control in the \awinfsys\ is integral to all
aspects of data handing and provides transparent access to quality estimators
for all stages of data reduction from the raw image to the final catalog.
The implementation of quality control mechanisms relies on the core features in
this \awenv (\awe): an object-oriented framework, full data lineage, and both forward and
backward chaining.  Quality control information can be accessed via the
command-line \aweprompt\ and the web-based Quality-WISE service.  The quality
control system is described and qualified using archive data from the 8-CCD Wide Field Imager (WFI)
instrument (\url{http://www.eso.org/lasilla/instruments/wfi/}) on the 2.2-m MPG/ESO telescope at La Silla
and (pre-)survey  data from
the 32-CCD OmegaCAM instrument (\href{http://www.astro-wise.org/~omegacam/}{http://www.astro-wise.org/$\sim$omegacam/}) on the VST telescope at Paranal.


\end{abstract}

\section{Introduction}\label{sec:intro}

Quality control is typically one of the greatest challenges in the chain from
raw sensor data to scientific papers.  This includes not only limited
observations for an individual scientist such as subsets of archival WFI data,
but also bulk observations of large astronomical surveys, such as those taken
with OmegaCAM on the VST (VLT Survey Telescope).  In such surveys, the human
and financial resources required often dictate that not only the large survey
teams are spread over many institutes in many countries, but also the required
data storage and the parallel computing resources.  Such a situation requires
an environment in which all non-manual qualifications are automated and the
scientist can graphically inspect where needed.  This is easily achieved by going back and forth 
through the data and metadata of the whole processing chain for large numbers of 
data products, and for only those data products where it is necessary.  Such 
efficiency is clearly as beneficial to individual scientists as it is to large 
survey teams. 

These requirements force survey teams beyond the era of science on a desktop
and dictate a paradigm in which astronomers, calibration scientists, and
computer scientists spread over geographically distant locations in many countries
share their work and latest results in a single environment that allows the
optimized processing, quality control, and archiving of large data sets.  This
means a federated system of humans, databases, computing resources, and data
storage yielding an \textit{integrated information system} \citep{adass}.
This integrated information system, \aw, is introduced and
described in detail in \citet{begeman}.  It is assumed that the reader is
familiar with the fundamental concepts described in these papers as only the
most relevant concepts will be dealt with here.

\subsection{Traditional quality control}

The quality control of astronomical data is a key to success in obtaining
necessary data for scientific use cases.  Quality control allows scientists to
verify observations, to improve observational plans, to correct the regime of
observations, to check the data processing and, finally, to distinguish between
an artifact and a real event detected during the observations.

Present day observations, especially the vast amounts in the case of large
astronomical surveys, require complicated processing systems involving a number
of data processing levels and programming efforts from many scientists and
programmers, usually distributed over a number of institutions.  Tracing
data quality through the processing chain given the involvement of many scientists 
and institutions becomes a non-trivial but crucial task.

There are many efforts invested in checking the quality of data delivered by an 
instrument, but this quality control remains at the observation/reduction site and 
comes to the scientific user as a reduced set of parameters describing the quality 
of the observations \citep{hummel,VISIR}.  There is no way for the user to return 
to the raw observational material and check the quality of a particular 
observation.  In the case when the user does not process the data her/himself, but 
accesses only the final product, she/he has to rely on the model of the quality 
control chosen by the people behind the data processing.  There is a general 
understanding that the quality control should be shared by the observers and 
scientists responsible for the data processing \citep{VLT}.  Nevertheless, this 
does not relieve the user from the task of making decision about the data quality
based on incomplete and non-reproducible information provided with the end product.

One mechanism bulk data providers employ to describe the quality of data products
is to introduce a number of attributes in the data model which will hold
information related to the quality control.  For example, in the case of 2MASS
data products the quality control was performed during the observations and the
data processing, and the final catalog was formed according to the algorithm
described in \citet{2MASS}.  From 60 attributes of the Two Micron All Sky Survey 
Point Source Catalog (2MASS PSC), 31 attributes are related to data quality.  This 
allows the user to create a subset according to his/her preferences for the 
quality of the data, but limits the user to the \textit{good quality} data.  The 
criteria for the data to be considered as \textit{good} are defined for a survey, 
not for a user of its data.  Similar approaches were used by SDSS and UKIDSS 
surveys.  In all these cases, data are delivered in a catalog with uniform quality 
rather than optimizing quality for particular data subsets \citep{SDSS,UKIDSS}.   
This is contrary to the typical goal of an individual scientist using the final 
data products.

To make a sound decision about data quality, the user should be able to access 
quality control algorithms at any point from the observation to the creation of 
the end product.  Thus, ideally, quality control should be performed on and 
reviewed at each processing step.  As a result, the user can trace the origin of 
any problem associated with quality parameters back to the specific processing 
step and/or the data entity responsible for it.

\subsection{\aw\ quality control}

\begin{figure}
\centering
\includegraphics[angle=0,width=118mm]{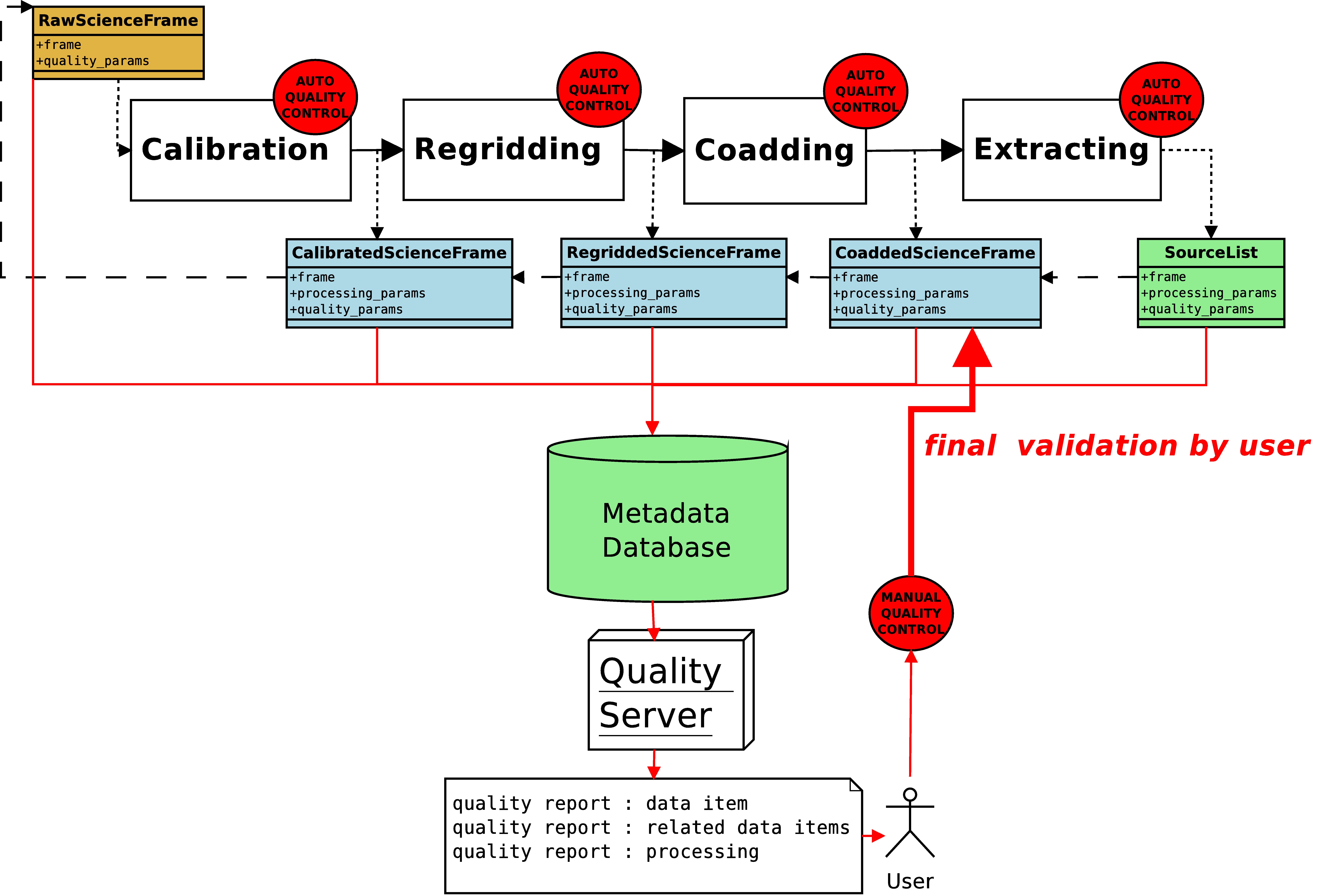}
\caption{
The quality control model in the \awinfsys.  Shown is a schematic of the
full data lineage with quality control at each processing step.
}\label{fig:quality}
\end{figure}

The core difference between this ``traditional'' quality control and the \aw\
approach to quality control is that the latter one uses features of \aw\ as an
\textit{integrated information system} to trace the quality at all stages of
data production.  These features are: data processing and quality control
within the same system, an object-oriented framework, and full data
lineage with both forward and backward chaining.  Together, they allow testing
of the quality of any data product, intermediate or final, from any other data
product at any stage of processing or analysis. The advantages to this approach
include allowing survey teams or individual scientists to inspect the quality of
any data product, allowing reprocessing of all or only part of one or multiple
data products in the most efficient way possible.  In this way, the user knows
exactly what the final quality means and can even reprocess any set of data to
her/his needs.

Figure~\ref{fig:quality} shows an integral approach of quality control supported 
by the \awinfsys.  There are two types of quality control at each stage of the 
data processing: automatic (default) and manual (optional).  The user can visually
inspect each data item and validate/invalidate it.  All the information about
the quality at every stage of data processing is saved in the database.

The object-oriented framework includes a set of parameters that are assigned
to each data class, and forms a built-in system of general quality estimators.  The
following section describes these quality parameters used in the
\awenv\ (\awe) and how they are connected between different types of data.
Section~\ref{sec:quali} describes the quality control mechanisms built into \awe.
Section~\ref{sec:trend} gives examples of how trends in any aspect of the data
can be isolated using the command-line (\aweprompt).  Finally,
Section~\ref{sec:qwise} describes the graphical interface for quality control in
\awe.

\section{Quality parameters}\label{sec:param}
\subsection{Data visibility}\label{sec:visib}

Visibility of data meeting the minimum level of quality to be processed
in \awe\ is governed by privilege level and by validity (i.e., privileged
data and data flagged as poor quality is hidden).   Privileges in \awe\
are levels of accessibility for different groups, similar to permissions
levels on a UNIX file system.

All data entities in \awe\ are instances of Object-Oriented Programming (OOP)
\textit{objects}.  Validity, and thus the processability, is
indicated by setting any or all of the following flag attributes of a given object:
\begin{enumerate}
\item
  \texttt{is\_valid} -- manual validity flag
\item
  \texttt{quality\_flags} -- automatic validity flag
\item
  \texttt{timestamp\_start/end} -- validity ranges in time (for calibrations only)
\item
  \texttt{creation\_date} -- the most recent valid data is the best
\end{enumerate}
For instance, obviously poor quality data can be flagged by setting its
\texttt{is\_valid} attribute to 0, preventing it from ever being processed
automatically.   The calibrations used are determined by their
\texttt{timestamp\_start}, \texttt{timestamp\_end}, and \texttt{creation\_date} attributes
(Which calibrations are valid for the given data?), and the quality of processed
data by the automatic setting of its \texttt{quality\_flag} attribute (Is the
given data good enough?).  Good quality data can then be flagged for promotion
(\texttt{is\_valid} $>1$) and eventually promoted in \textit{privilege} by its
creator (published from level 1 to 2) so it can be seen by the project manager
who will decide if it is worthy to be promoted once again (published from level
2 to 3 or higher) to be seen by the greater community.  In the end, publishing
of data and results can be done by the manual setting of a single flag
attribute\footnote{All of these attributes can be modified via the command-line
\aweprompt\ or via one or more web services (see Sect.~\ref{sec:qwise}).}.

The example below shows how the user can invalidate a particular bias frame for
a particular instrument, detector and date using \awe.
\begin{verbatim}
awe> bias = BiasFrame.select(instrument='WFI', chip='ccd57',
....                         date='2003-10-05')
awe> print bias.is_valid
1
awe> context.update_is_valid(bias,0)
awe> print bias.is_valid
0
\end{verbatim}
Note that the query returns the most recent, valid master bias object for the
given criteria.  This same mechanism is used to query for objects
during processing.

\subsection{Provenance: full dependency linking and data lineage}\label{sec:prove}

The \awenv\ uses its federated database \citep{begeman,adass} to link all data products to their
progenitors (dependencies), creating a full data lineage of the entire
processing chain.  This allows quick and simple troubleshooting of data results
by looking at processing settings, calibrations and more.  It also allows for
direct monitoring of the progress of survey or individual observations, thus
simplifying observation management.  This data lineage
also provides the ability to analyze trends in dependencies to aid in
troubleshooting (see Sect.~\ref{sec:fivel}).


Raw data is linked to the final data product via database links within the
\textit{data object}, allowing all information about any piece of data to be
accessed instantly.  See \citet{mwebaze} for a detailed description of \awe's
data lineage implementation.  This data linking uses the power of OOP to
create this framework in a natural and transparent way.

\section{Built-in quality control mechanisms}\label{sec:quali}

In the \awenv, quality control permeates all aspects of the data reduction process.
From the moment data enters the system, through all processing steps, to the
final data product, data quality is retained and can be accessed transparently.
This is accomplished by integrating quality control concepts at the lowest
levels of the system.

\subsection{Integrated quality control}\label{sec:integ}

Quality control of the reduction process in \awe\ is integrated directly into
the objects.  Three methods exist on all {\proce}s (the afore mentioned OOP
\textit{objects} that describe data entities undergoing some level of
processing):

\begin{itemize}
\item{\texttt{verify()}}
compares values derived from the current \proce\ instance to known acceptable
limits (e.g., image statistics) and automatically raises
\texttt{quality\_flags} if the limits are exceeded
\item{\texttt{compare()}}
compares values derived from the current \proce\ instance to those of the
previous version and automatically raises \linebreak \texttt{quality\_flags} if 
the values are worse
\item{\texttt{inspect()}}
provides an interface for manual inspection of the current \linebreak \proce\ 
instance (e.g., viewing the image pixels)
\end{itemize}

The quality control parameters are stored in two persistent properties of the
object, \texttt{is\_valid} and \texttt{quality\_flags}.  As mentioned before,
the \texttt{is\_valid} property is the manual flag used to validate or
invalidate any \proce, and the \texttt{quality\_flags} property stores the
results of the automatic verification routines.  This model shares similarities
with other quality control ``scoring'' models (e.g., \citet{hanuschik}) and is discussed in the processing context in Sect.~\ref{sec:proce}.

To give examples in contrast to this model, the Sloan survey uses automated 
pipelines (e.g., \textit{runQA} and \textit{matchQA}) run separately from the 
processing pipeline to assess and report the quality of the data \citep{SDSS}, and 
the UKIDSS survey employs the metadata storage of FITS images to convey quality 
parameters to the QC procedures (\citet{UKIDSS} and reference D06 therein).  The 
integrated nature of the quality parameters and procedures in \awe\ has clear 
advantages over these other models because the quality parameters are directly 
part of the \proce.

This integrated quality control is one of the simplest, yet most powerful
aspects of \awe\ for survey operator and individual scientist alike.  Both high
and low quality data can be accessed via a simple query and the cause of the
low quality can be known directly via the bit-masked value of its
\texttt{quality\_flags} attribute.  Also, the nature of the queries in the
processing recipes guarantees that low quality data is never processed unless it
is manually specified.

This paradigm for quality control allows for construction of tools such as
\qwise\footnote{\url{http://quality.astro-wise.org/}} that can act as the
QC front-end of the entire system.  Data quality (of both pixel data and its
metadata) can be viewed through a simple interface.  This interface allows access to flagging of
data (triggering automatic reprocessing), to direct reprocessing of data and
even to the quality of linked objects.  This all exists within the information
system allowing effective sharing of human resources.

\begin{table}[htb]
\begin{center}
\begin{tabular}{|l|l|r|r|}
\hline
\textbf{Class} & \textbf{process\_param}     & \textbf{value} & \textbf{units}\\
\hline
\rawbi & max\_bias\_stdev                    &         100.0  & ADU       \\
       & max\_bias\_level                    &         500.0  & ADU       \\
       & max\_bias\_flatness                 &          10.0  & ADU       \\
\hline
\rawdo & min\_flat\_mean                     &        5000    & ADU       \\
       & max\_flat\_mean                     &       55000    & ADU       \\
\hline
\rawtw & min\_flat\_mean                     &        5000    & ADU       \\
       & max\_flat\_mean                     &       55000    & ADU       \\
\hline
\readn & maximum\_readnoise                  &           5.0  & ADU       \\
       & maximum\_bias\_difference           &           1.0  & ADU       \\
       & maximum\_readnoise\_difference      &           0.5  & ADU       \\
\hline
\gainl & maximum\_gain\_difference           &           0.1  & $e^-/$ADU \\
       & minimum\_gain                       &           2.0  & $e^-/$ADU \\
       & maximum\_gain                       &           5.0  & $e^-/$ADU \\
\hline
\biasf & maximum\_stdev                      &          10.0  & ADU       \\
       & maximum\_stdev\_differ              &          10.0  & ADU       \\
       & maximum\_subwin\_flatn              &      100000.0  & ADU       \\
       & maximum\_subwin\_stdev              &      100000.0  & ADU       \\
\hline
\hotpi & maximum\_hotpixelcount              &       50000    &           \\
       & maximum\_hotpixelcount\_difference  &         100    &           \\
\hline
\coldp & maximum\_coldpixelcount             &       80000    &           \\
       & maximum\_coldpixelcount\_difference &         100    &           \\
\hline
\domef & maximum\_subwin\_flatness           &        1000.0  & ADU       \\
       & maximum\_subwin\_diff               &        1000.0  & ADU       \\
\hline
\twili & maximum\_subwin\_flatness           &        1000.0  & ADU       \\
       & maximum\_subwin\_diff               &        1000.0  & ADU       \\
       & maximum\_number\_of\_outliers       &       10000    & ADU       \\
\hline
\maste & maximum\_subwin\_diff               &        1000.0  & ADU       \\
\hline
\photo & max\_error                          &           0.03 & mag       \\
\hline
\astro & min\_nref                           &          15    &           \\
       & max\_nref                           &        1200    &           \\
       & max\_sigma                          &           1.0  & arcsec    \\
       & max\_rms                            &           1.0  & arcsec    \\
       & min\_n\_overlap                     &          20    &           \\
       & max\_n\_overlap                     &       20000    &           \\
       & max\_sigma\_overlap                 &           0.1  & arcsec    \\
       & max\_rms\_overlap                   &           0.1  & arcsec    \\
\hline
\end{tabular}
\end{center}
\caption{
Representative examples of QC limits used by the automated \texttt{verify()}
and \texttt{compare()} methods on the given class instances (objects).  These
examples are limited to calibration data and are derived from the requirements
for the OmegaCAM instrument and updated based on experience with archive data of
the WFI instrument.  See the document page linked from the class name of
appropriate links on \url{http://doc.astro-wise.org/astro.main.html} for more 
details.
}\label{tab:proce}
\end{table}

\subsection{Quality control during ingestion}

A number of automatic, simple quality control procedures are executed at
the lowest level of data interaction--ingestion into the system.  These
procedures are used to flag poor-quality data so they are excluded from further use.
The procedures include checks on the median and standard deviation of the
pixel values in bias exposures, and the exposure level of flat-fields.
The levels at which flags are raised are instrument and detector chip
dependent, as needed.

\subsection{Quality control during processing}\label{sec:proce}

Quality control at the processing stage starts well before any actual
processing is done.  The selection of data to be processed is subject to the
visibility mechanism (see Sect.~\ref{sec:visib}).  All processing tasks
first check the validity and quality of candidate science data, and the
validity, quality and timestamp ranges of applicable calibration data.
This guarantees that only the highest quality data is considered for
processing.

Once data processing is complete, the quality methods of data product object are
run to verify that this is the highest quality product
possible (see Sect.~\ref{sec:integ}).  The \texttt{verify()} and \texttt{compare()} methods are
automatically run to check the data product against the accepted limits and to
make sure the quality is higher than the previous version if one exists.  If
either test fails, one or more \texttt{quality\_flags} are raised.
Table~\ref{tab:proce} gives a representative sample of the limits tested via the
\texttt{verify()} and \texttt{compare()} methods.  Optionally, the
\texttt{inspect()} method can be run manually to interactively check the data
product.  A non-interactive version of this method is always run to create and
store a static version of the inspection plot for later perusal via the
command-line or through the \qwise\ service (see Sect.~\ref{sec:qwise}).

\subsection{Inspection plots}\label{sec:inspe}

During processing, quality control inspection plots are made as a matter of
course.  These can be viewed interactively during processing or saved for later
viewing.  As most processing is done in a parallel environment, these
inspection plots tend to have a very low creation cost.

\begin{figure}
\begin{center}
\includegraphics[angle=0,width=45mm]{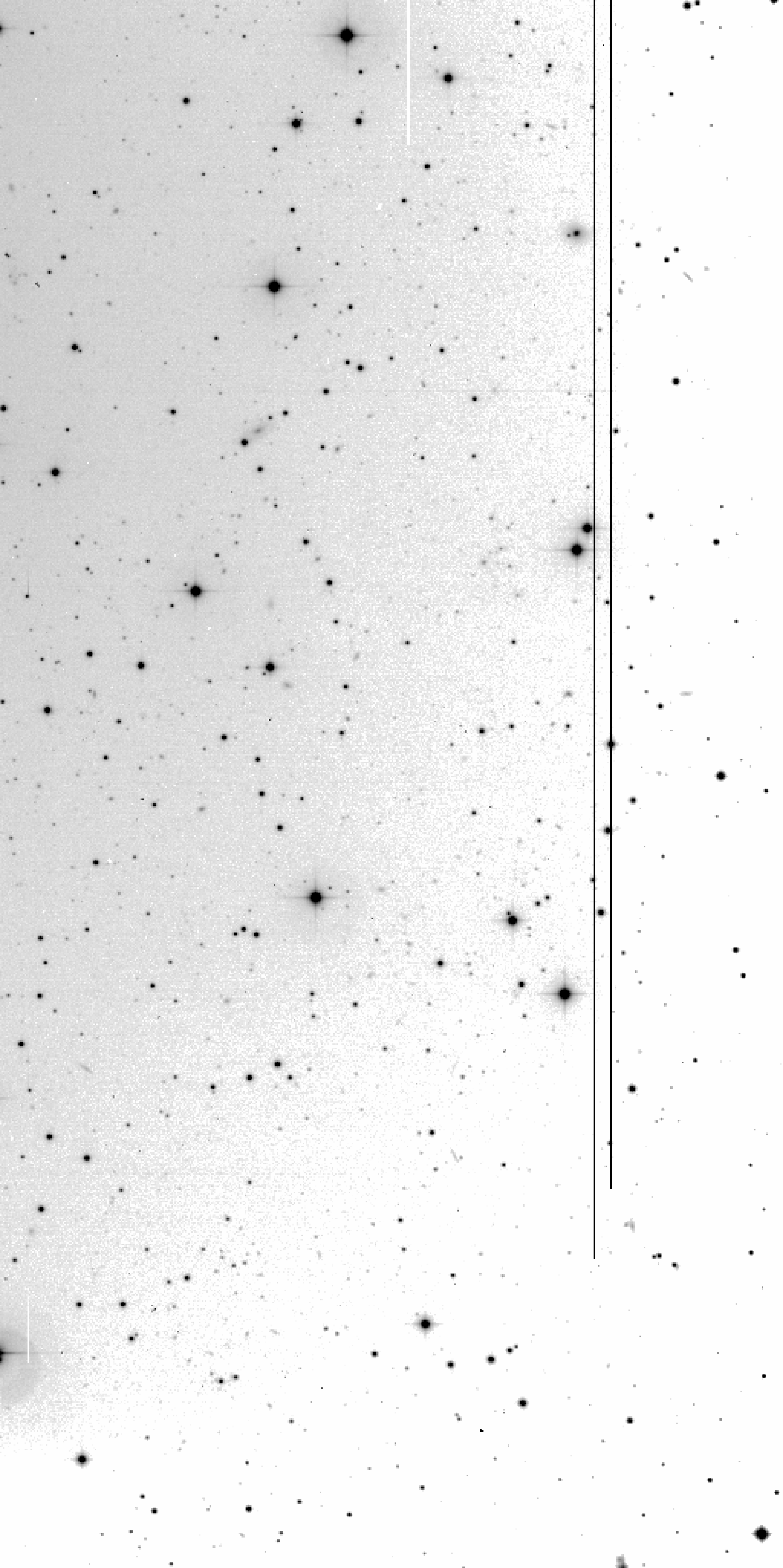}
\hspace{8mm}
\includegraphics[angle=0,width=45mm]{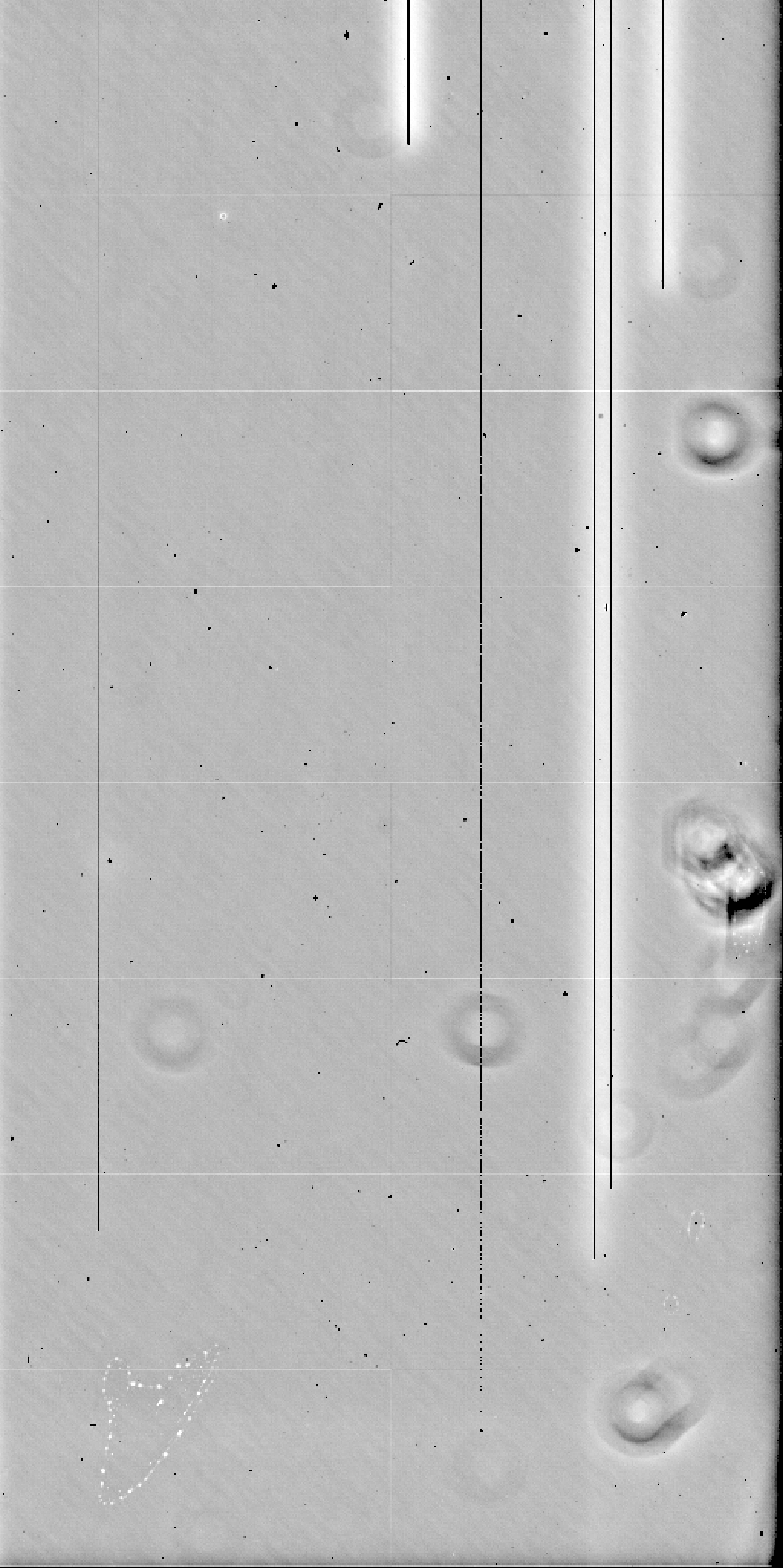}
\caption{
(left panel) A thumbnail representation of a WFI \reduc\ created by STIFF.  The optimized
intensity cuts and binning allow a quick assessment of the quality.  This
particular example shows an intensity gradient caused by either poor flat
fielding, nebulosity from a galaxy at the center of the mosaic field (to the
upper left), or simply a non-uniform illumination of the focal plane.  The
intensity values are inverted.
(right panel) A thumbnail representation of a WFI \weigh\ created by STIFF.  The optimized
intensity cuts and binning allow a quick assessment of the quality.  This
particular example is associated with the thumbnail in the left panel.
Saturated stellar peaks and bad columns are clearly visible in addition to
``doughnuts'' of the primary mirror of the telescope that are part of the flat
field foundation of the \weigh.  White pixels have values near 1, black pixels
have values at or near 0.  The horizontal lines are artifacts of the CCD
manufacturing process.  The higher weight of the pixels near some of the bad
columns is an artifact caused by Fourier processing of input flat frames
without properly taking into account bad pixels.  It is possible to identify
some of these defects with pixel statistics \textit{a priori}, but these
unusual cases are generally only identified through this type of inspection plot.
}\label{fig:ithum}
\end{center}
\end{figure}

\begin{sidewaysfigure}
\begin{center}
\includegraphics[angle=270,width=165mm]{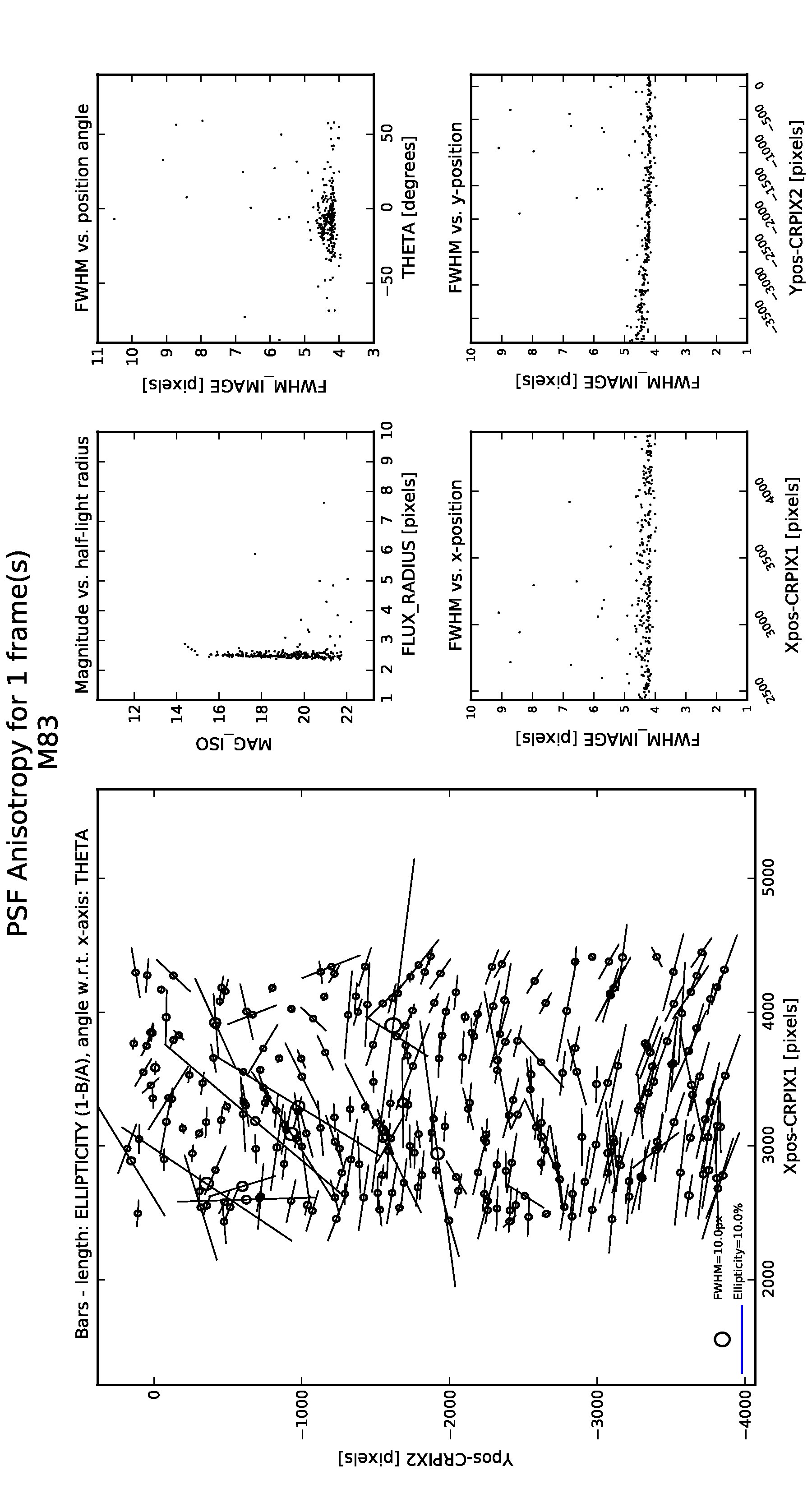}
\caption{
A PSF Anisotropy plot of the \reduc\ whose thumbnail is seen in
Fig.~\ref{fig:ithum}.
The left panel
shows the two-dimensional anisotropy in the PSF (in both FWHM and in
ellipticity) of the sources.  The top-middle panel shows source magnitudes
on an arbitrary scale versus flux radius and gives an indication of how
``stellar'' the sample is.  The remaining panels plot FWHM of the sources
versus ellipticity position angle, horizontal position and vertical
position, respectively, clockwise from top-right.
}\label{fig:ipsfa}
\end{center}
\end{sidewaysfigure}

\begin{figure}
\begin{center}
\includegraphics[angle=0,width=80mm]{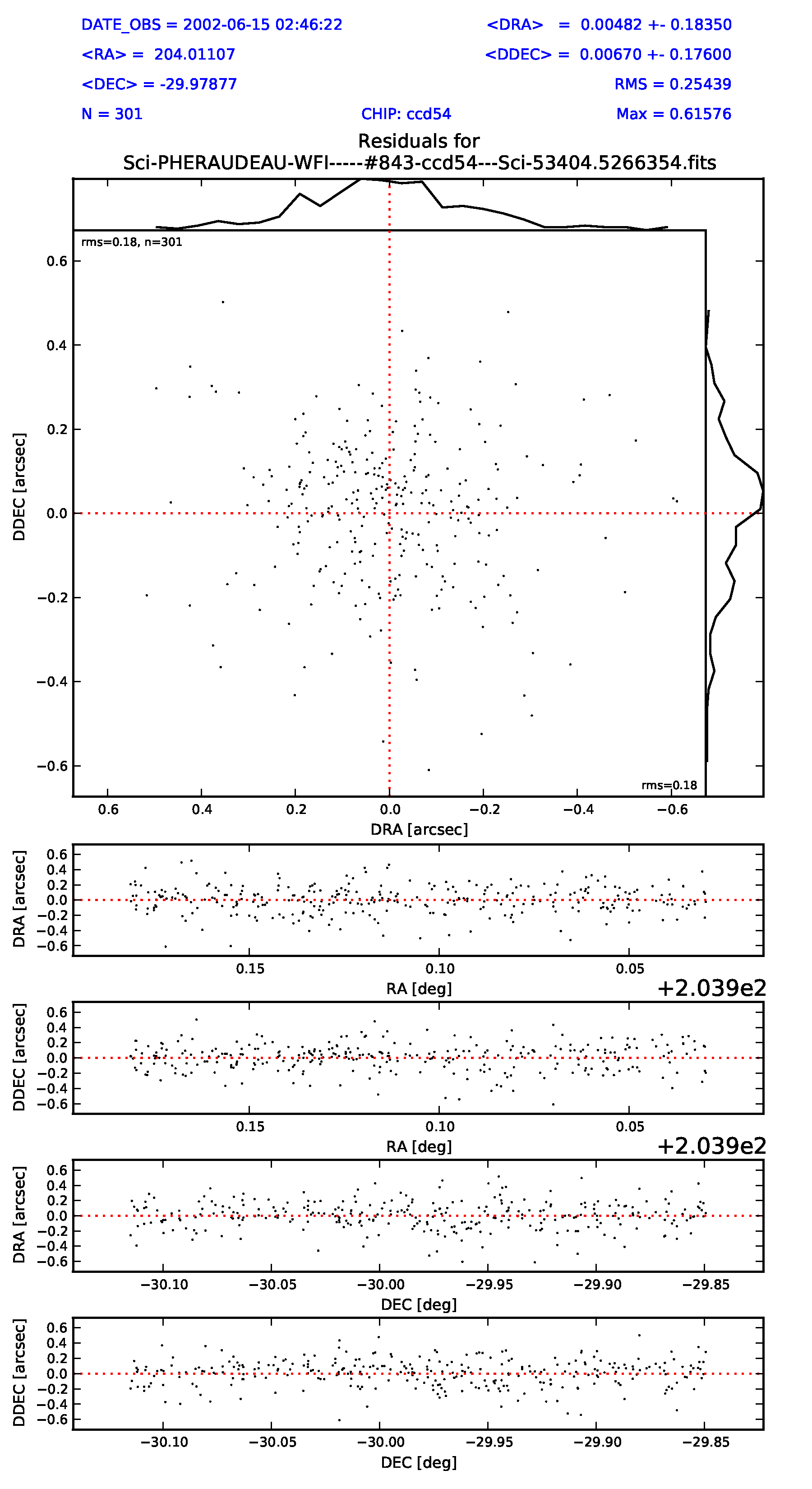}
\caption{
An \astro\ inspection plot for a recent solution of the \reduc\
associated with the thumbnail in Fig.~\ref{fig:ithum}.  The plot displays the
statistics of the residuals (DRA and DDEC) between the RA and DEC of sources in
a source catalog to which the local astrometric solution has been applied and
the RA and DEC of those sources as listed in the reference catalog of
astrometric standards, USNO-A2.0 in this case.  The text in the top of the figure lists the observation
date (DATE\_OBS), the number (N) of sources pairs plotted, their average RA
($<$RA$>$) and DEC ($<$DEC$>$) in degrees, the average RA and DEC residuals
($<$DRA$>$ and $<$DDEC$>$) and their standard deviations in arcsec, and finally
the root-mean-square (RMS) of the two-dimensional residual and the maximum
two-dimensional residual (Max) in arcsec.  The large upper panel plots DRA
versus DDEC.  The four panels below it show DRA and DDEC with respect to RA
(with a constant offset of 203.9 degrees) and then to DEC.
}\label{fig:iastr}
\end{center}
\end{figure}

\begin{figure}
\begin{center}
\includegraphics[angle=0,width=90mm]{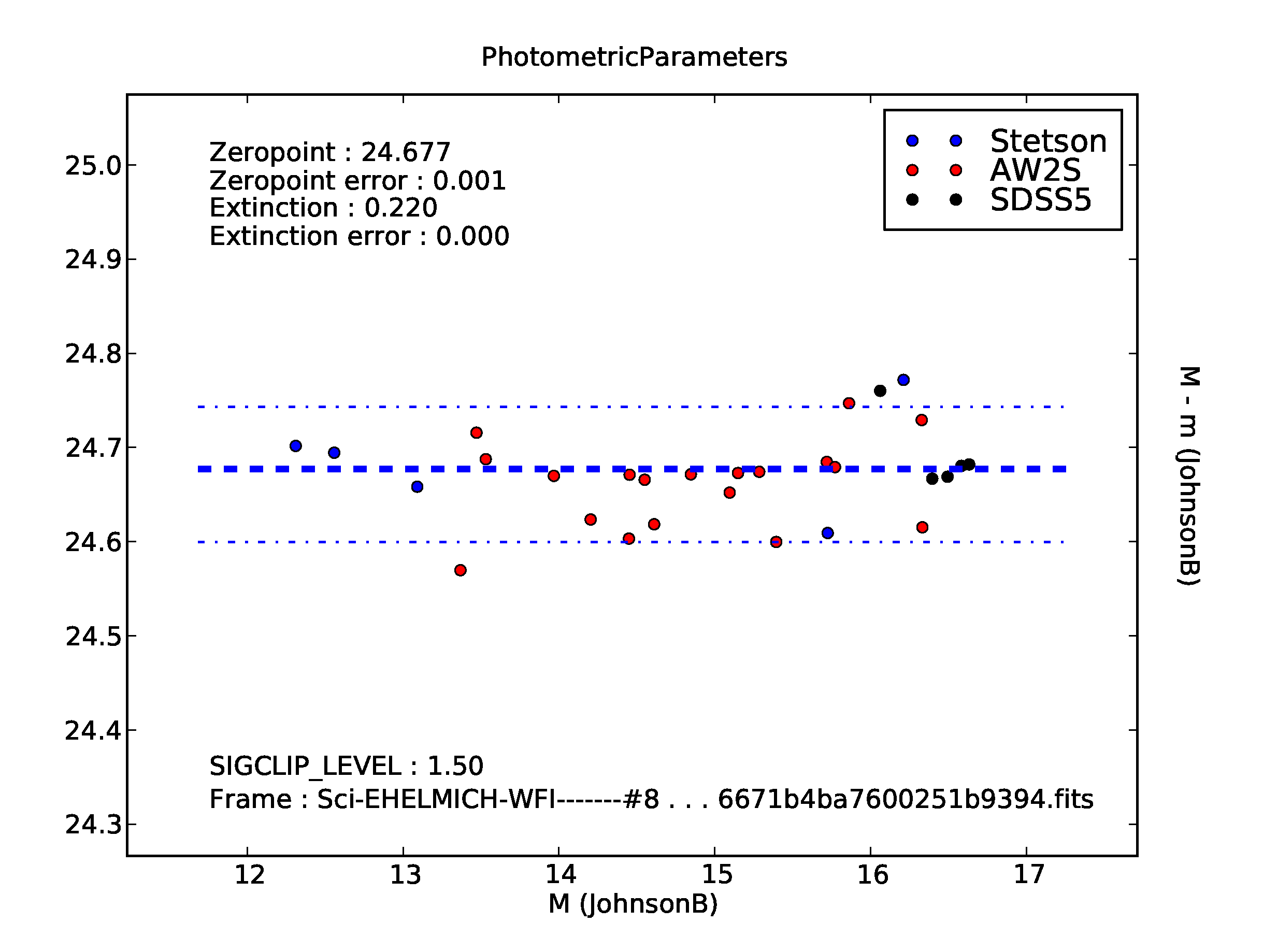}
\caption{
A \photo\ inspection plot for a photometric observation comprising one WFI
detector.  A graphical representation of the data used to calculate the
photometric zeropoint.  In this plot, three photometric reference catalogs can
be seen: Stetson (blue points), \aw\ secondary standards (red points) and Sloan
Digital Sky Survey data release 5 (black points).
}\label{fig:iphot}
\end{center}
\end{figure}

\begin{figure}
\begin{center}
\includegraphics[angle=0,width=90mm]{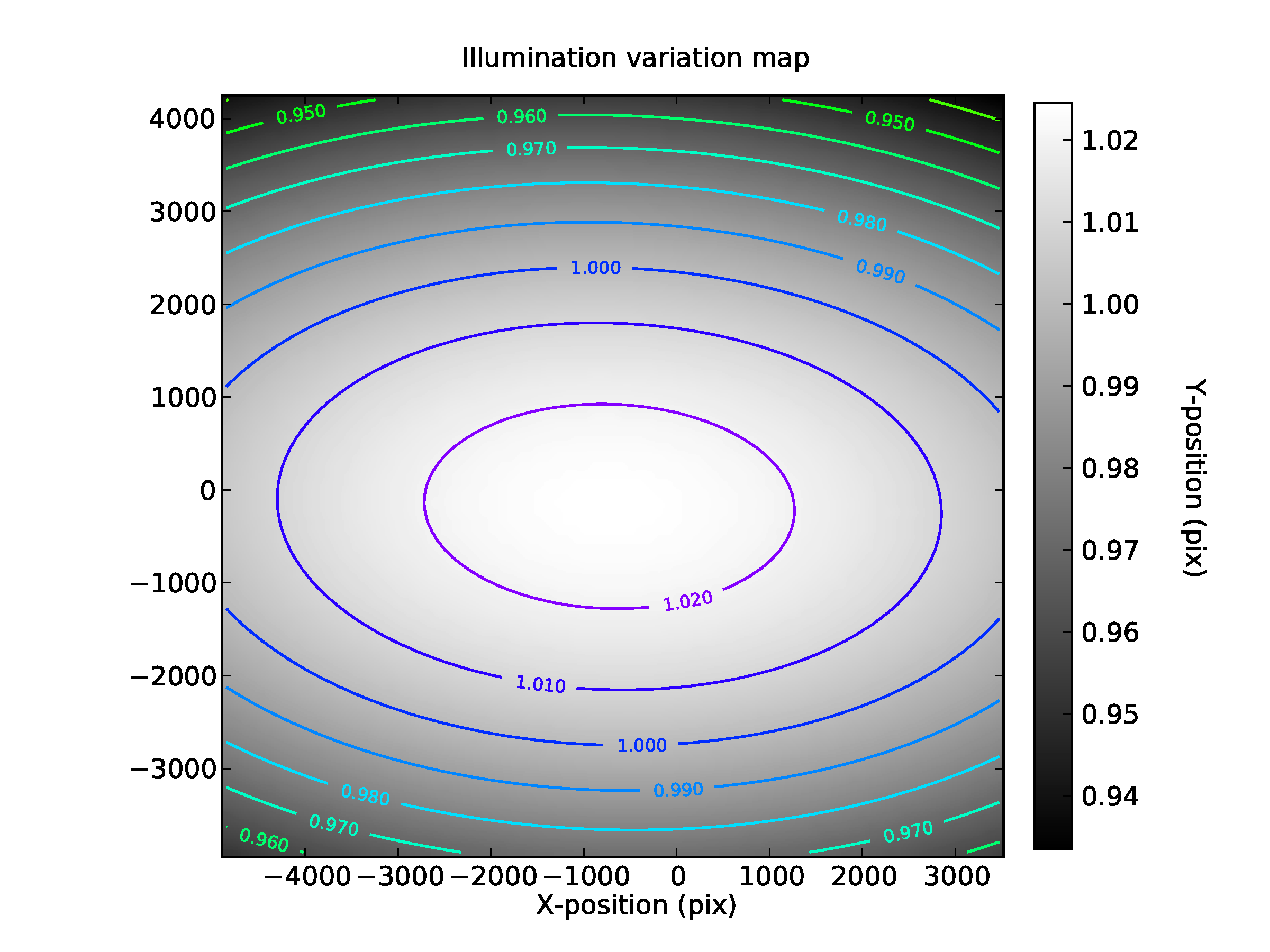}
\caption{
An \texttt{IlluminationCorrection} inspection plot for WFI data.  This plot is
a schematic representation of the illumination variations across the region of
consideration, usually the entire field-of-view, eight 2k$\times$4k detectors
in this case.
}\label{fig:iillu}
\end{center}
\end{figure}

Inspection plots exist for many of the object types in \awe, particularly
those critical for assessing the quality of major data products (e.g., science
data quality, end-to-end detrending quality, astrometric and photometric
calibration quality).  See Fig.~\ref{fig:ithum}~through~\ref{fig:iillu} for
examples of such plots.

These static plots are simple snapshots of the most useful information to be
inspected.  In \awe, there exists the ability in most cases to interact with
the inspection plot.  This is done using the PyLab interface to MatPlotLib.
This interface is integrated into \awe, and forms the backbone of all types of
plotting, including post-processing analysis.

\section{Trend analysis}\label{sec:trend}

Many powerful ways exist in the \awenv\ to examine both pixel data and
metadata.  One of these ways is through the use of the command-line interface,
the \aweprompt.  Through this interface, one can examine individual quality
parameters and processing parameters of any object or linked object
transparently.

\subsection{Five-line script}\label{sec:fivel}

\awe\ consists of Python classes representing {\proce}s that can be created by
scripts (called recipes or Tasks).  The Tasks are simply sophisticated versions
of what are termed five-line scripts\footnote{The term file-line script derives
from the observation that most simple tasks in \awe\ can be achieved in about
five lines of code.} (5LS).  It is these 5LSs that do the bulk of the work of
the data reduction and analysis for the user.  The 5LS is also a powerful tool
for quality control as atypical objects can be isolated easily.

This 5LS concept is a very simple and powerful way for users to interact with
the data contained in the system.  They can be ``one-off'', ``on-the-fly'', or
``throw-away'' scripts used to locate some interesting aspect of the data, can
be written down in a source file for potential use at a later time, or can be
integrated into an existing or future Task for the benefit of the system.  One
set of examples of 5LSs focuses on seeing how aspects of raw data in the system
change over time, another gathers statistical data for comparison and outlier
detection, and the last quickly investigates a scientific aspect of existing
data in the system.

\begin{figure}
\begin{center}
\includegraphics[angle=0,width=90mm]{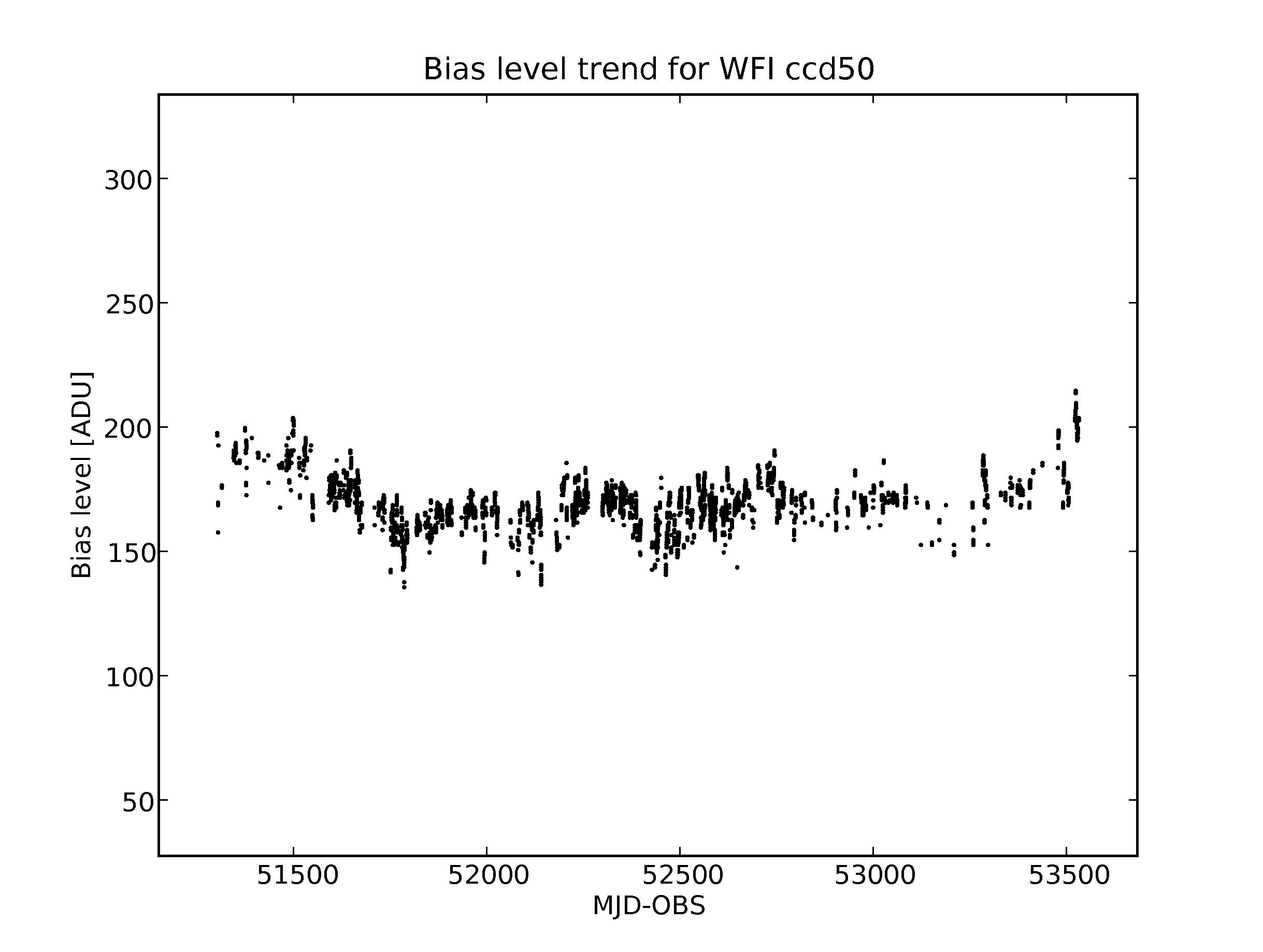}
\caption{
Plot of the bias level (median value of the science region) of ccd50 of the WFI
instrument from May 1999 to June 2005.
}\label{fig:biast}
\end{center}
\end{figure}

\subsection{Bias levels}

Display the bias level as a function of time for chip \textsf{ccd50} of the WFI camera:
\begin{verbatim}
awe> q = (RawBiasFrame.chip.name == 'ccd50') &\
....     (RawBiasFrame.quality_flags == 0) &\
....     (RawBiasFrame.is_valid > 0)
....
awe> biases = list(q)  # instantiate all biases
awe> x = [b.MJD_OBS for b in biases]
awe> y = [b.imstat.median for b in biases]
awe> pylab.scatter(x,y,s=0.5)
\end{verbatim}
This script will result in a plot similar to that seen in Fig.~\ref{fig:biast}.
It is important to note how the query is done.  Not only are the
objects of the desired detector queried for, the quality and validity (see
Sect.~\ref{sec:visib}) are also checked.  This prevents any data that are out of
specified ranges from being plotted, thus removing the worst outliers in the
resulting plot before the data is even compiled.  This lends significant
efficiency to this method of visualization.

\begin{figure}
\begin{center}
\includegraphics[angle=0,width=90mm]{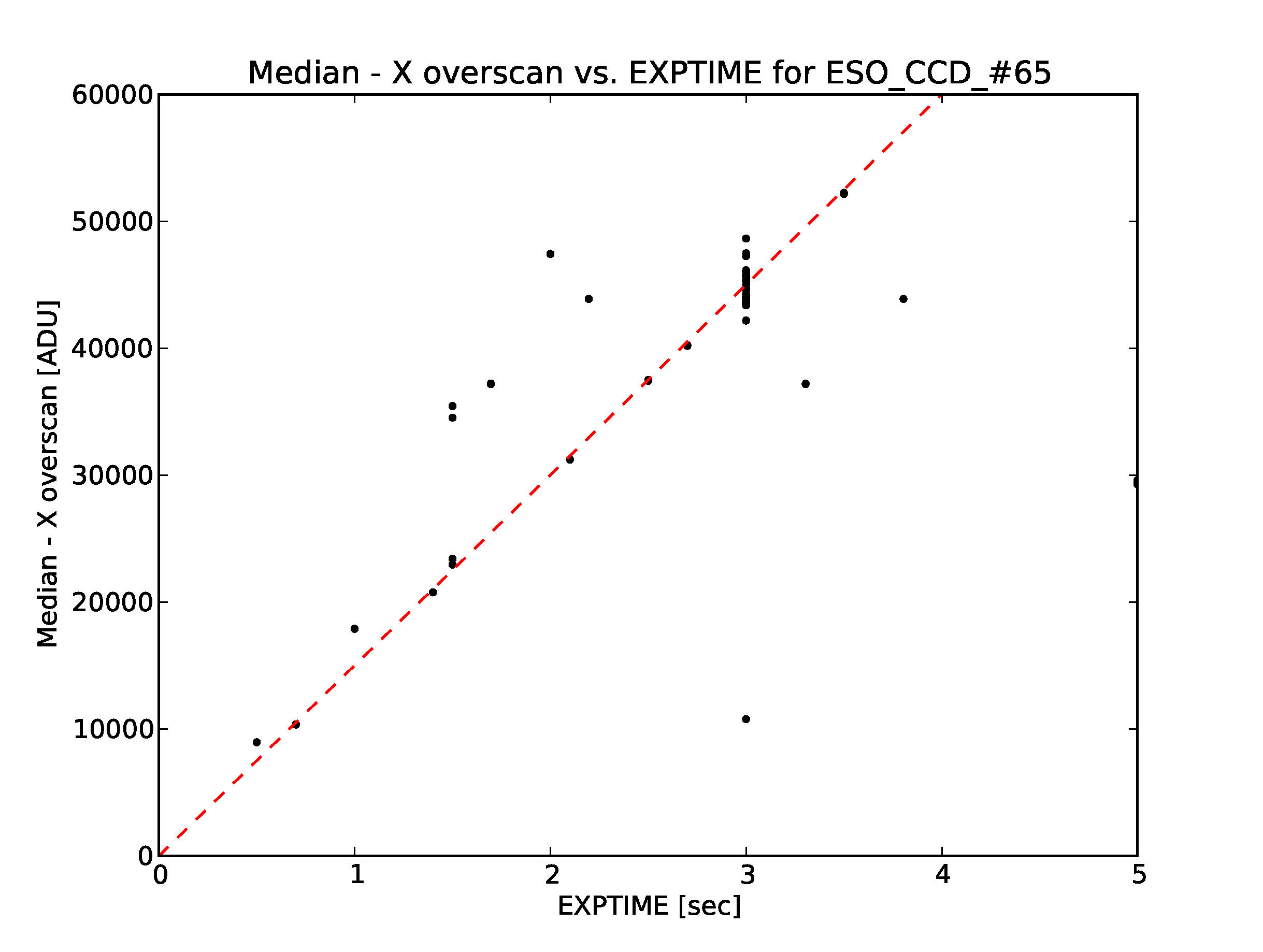}
\caption{
Plot of the dome flat exposure level (median value of the science region minus
the median value of the X overscan region) versus exposure time of
ESO\_CCD\_\#65 of the OmegaCAM instrument from data taken in 2011.  This plot
gives a quick indication of how linear this detector is.  The dashed red line
is only an indication of the slope in the data.  The cluster of points at 
EXPTIME$=$3 sec is from heavier sampling for diagnostic and detector health 
purposes.
}\label{fig:media}
\end{center}
\end{figure}

\subsection{Exposure levels}

\begin{figure}
\begin{center}
\includegraphics[angle=0,width=90mm]{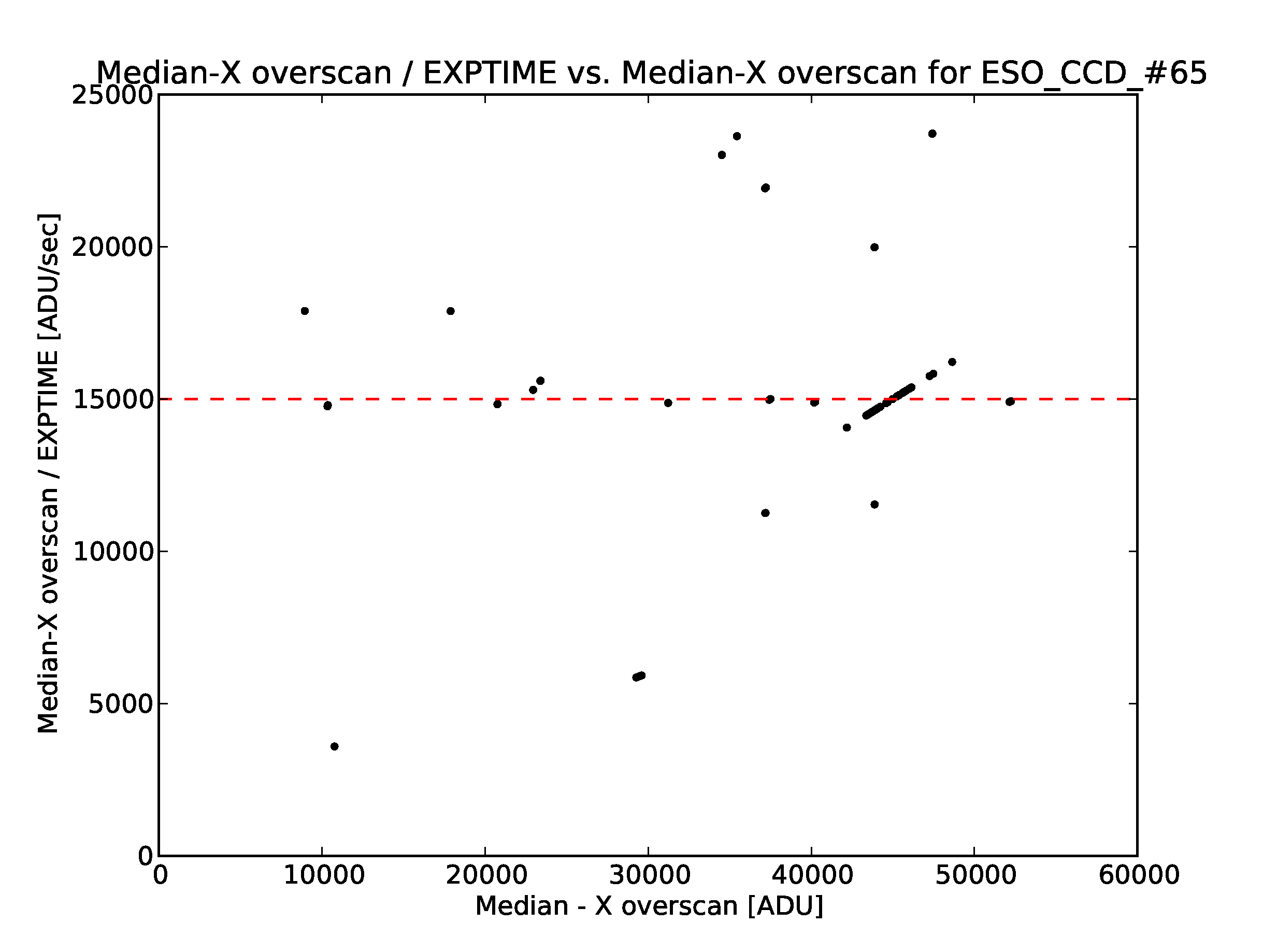}
\caption{
Plot of the dome flat exposure (median value of the science region minus the
median value of the X overscan region divided by the exposure time) versus the
dome flat exposure level (median value of the science region minus the median
value of the X overscan region) of ESO\_CCD\_\#65 of the OmegaCAM instrument
from data taken in 2011.  This plot quickly gives a different view of how
linear this detector is.  The dashed red line is only an indication of the
mean detector exposure.
}\label{fig:expos}
\end{center}
\end{figure}

Not only can simple values be plotted over time as in the previous section, but more complex
investigations of object attributes can be performed easily.  In this set of
examples, the linearity of an OmegaCAM detector is investigated:
\begin{verbatim}
awe> q = list((RawDomeFlatFrame.chip.name == 'ESO_CCD_#65') &
....          (RawDomeFlatFrame.filter.name == 'OCAM_g_SDSS') &
....          (RawDomeFlatFrame.quality_flags == 0) &
....          (RawDomeFlatFrame.is_valid > 0))
....
awe> exptime = [d.EXPTIME for d in q]
awe> med = [d.imstat.median-d.overscan_x_stat.median for d in q]
awe> pylab.plot(exptime, med, 'k.')
awe> pylab.plot([0,4], [0,60000], 'r--')
\end{verbatim}
This first example gives a plot similar to that shown in Fig.~\ref{fig:media}.
It is the overscan-corrected counts compared to the exposure time for
one detector of the OmegaCAM mosaic.  Simple arithmetic is seen in the
\textit{list comprehension} that creates the \texttt{med} list.  The second
example uses the data from the first, but adds the ability to perform array
arithmetic using NumPy\footnote{\url{http://numpy.scipy.org/}} to plot the
desired result (Fig.~\ref{fig:expos}).
\begin{verbatim}
awe> med = numpy.array(med)
awe> exptime = numpy.array(exptime)
awe> pylab.plot(med, med/exptime, 'k.')
awe> pylab.plot([0,60000], [15000,15000], 'r--')
\end{verbatim}

\noindent
This second example gives a quick exposure time-independent view of the same
data.  As in the result of the previous script, outliers can easily be seen.
It is now easy to isolate these outliers with NumPy methods using visually
chosen limits:
\begin{verbatim}
awe> outlier_mask  = (med/exptime < 10000)
awe> outlier_mask |= (med/exptime > 20000)
awe> outliers = med[outlier_mask], exptime[outlier_mask]
awe> good_data = med[~outlier_mask], exptime[~outlier_mask]
\end{verbatim}

\subsection{Twenty thousand light curves}

In the Fall of 2006, an investigation of light curves of the stars in the region of Centaurus-A\footnote{See \url{http://www.astro-wise.org/Presentations/LCnov06/CenA\_5LS\_valentijn/ for the details of the investigation and the various scripts used.}} was undertaken using
pre-reduced data in the \aw\ system.
The data was originally observed in the first half of 2005 with the WFI
instrument.  Only example scripts and resulting plots are reproduced here.  The
scripts have been updated and reformatted for inclusion.

\begin{figure}
\begin{center}
\includegraphics[angle=0,width=90mm]{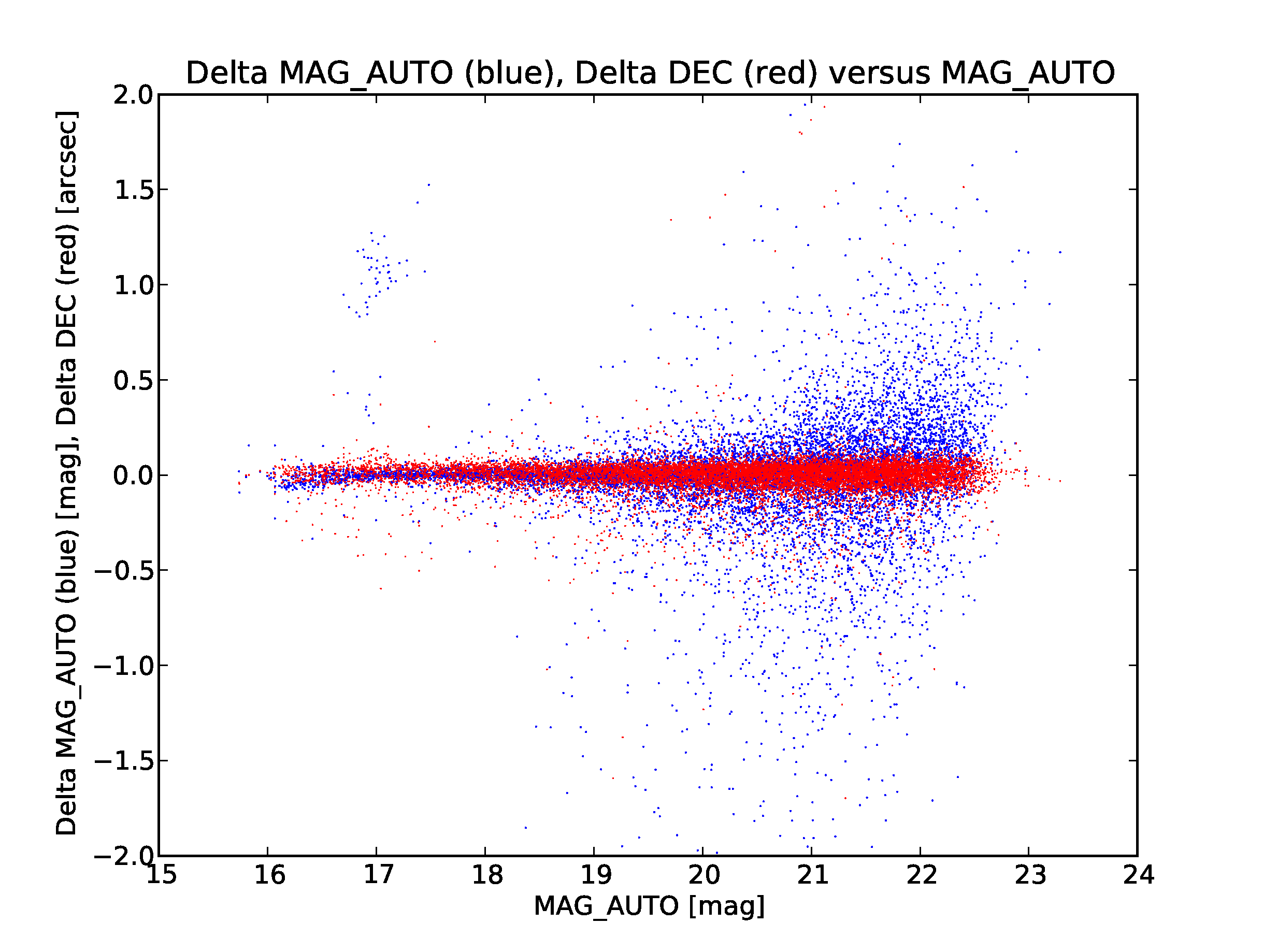}
\caption{
A plot of delta MAG\_AUTO (blue points) over-plotted with delta DEC (red points)
versus MAG\_AUTO.  The increase in scatter of the astrometric residuals is far
lower than that of the photometric residual, a qualitative indication that
astrometry for faint sources is at acceptable levels.
}\label{fig:delta}
\end{center}
\end{figure}

The first example takes data from an association of two coadded frames.  These
data exist in the system as an \assoc\ object.  Some astrometric and
photometric parameters are \textit{mined} from the association data.  This is
plotted in such a way to test the astrometric accuracy of fainter sources (see
Fig.~\ref{fig:delta}).  The plot clearly shows a slight degradation in this
accuracy, but also shows that it is not a source of concern as the position of
faintest sources is still generally well known.
\begin{verbatim}
awe> Al = (AssociateList.ALID == 1431)[0]
awe> arlist =  ['RA', 'DEC', 'MAG_ISO', 'MAG_AUTO', 'MAG_APER']
awe> r = Al.get_data_on_associates(arlist,mask=3,mode='ALL')
awe> mag, dmag, ddec = [], [], []
awe> for aid in r.keys():
....     # index 0 = SLID, 1 = SID, # added automatically
....     # index 3 = DEC, 5 = MAG_AUTO
....     mag.append(r[aid][0][5])
....     dmag.append(r[aid][0][5] - r[aid][1][5])
....     ddec.append((r[aid][0][3] - r[aid][1][3])*3600)
....
awe> pylab.plot(mag, dmag, 'b.', ms=0.5)
awe> pylab.plot(mag, ddec, 'r.', ms=0.2)
awe> pylab.ylim([-2,2])
\end{verbatim}

\begin{figure}
\begin{center}
\includegraphics[angle=0,width=90mm]{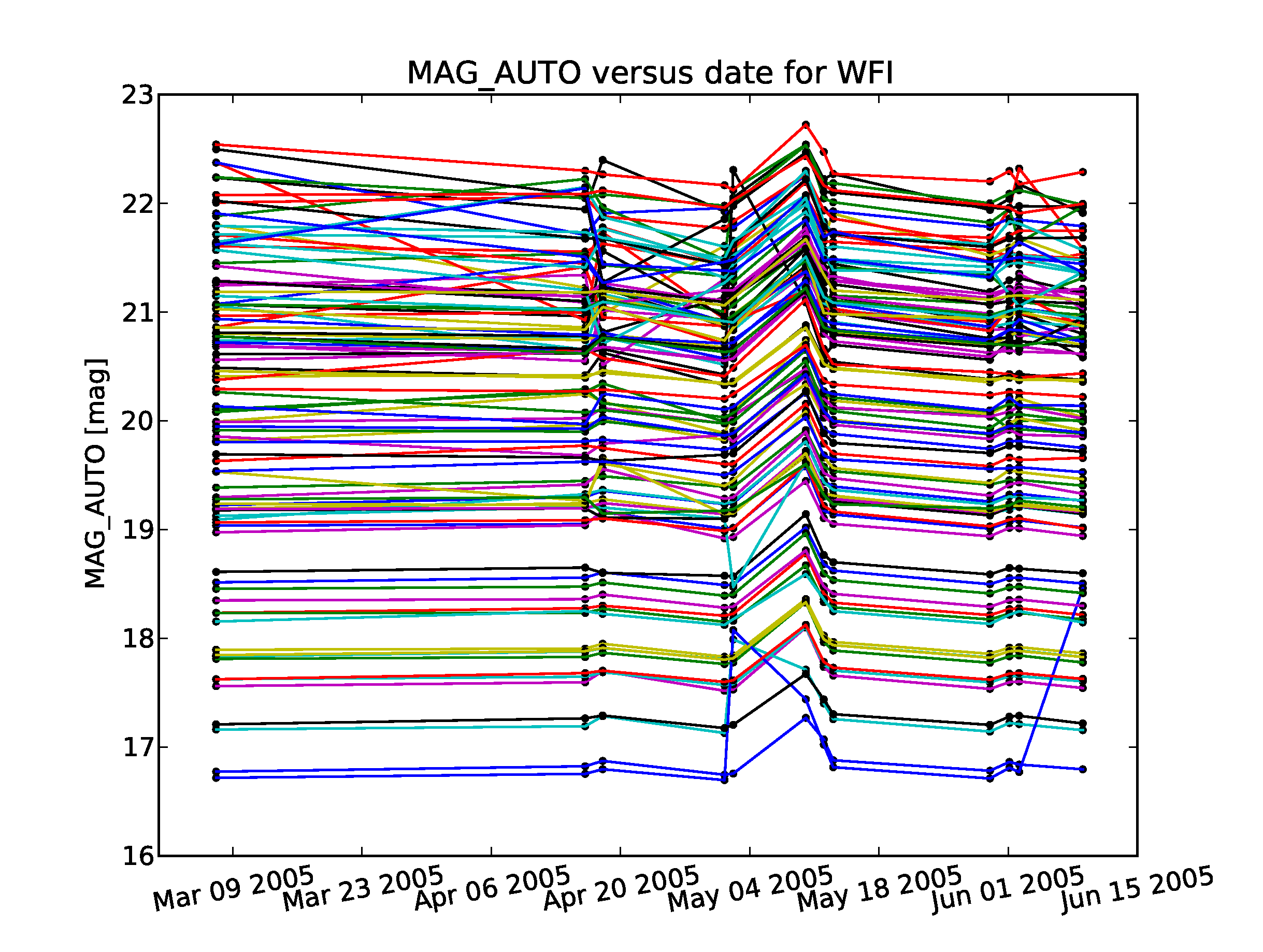}
\caption{
A plot of MAG\_AUTO versus the date for 100 of approximately 7500 light-curves
containing 12 photometric data points.  It is obvious that there remain
systematic offsets in the zeropoints.
}\label{fig:light}
\end{center}
\end{figure}

The next example mines data and creates a plot of light curves for
approximately 7500 of the 20000 stars associated with at least one other star
in one of the other observations.  These 7500 are the stars that were
associated for all 12 observations (i.e., where photometric data exists for all
12 observations).  For brevity and clarity, only the first 100 of these are
plotted by the script and shown in the accompanying plot (see
Fig.~\ref{fig:light}).
\begin{verbatim}
awe> Al = (AssociateList.ALID == 1534)[0]
awe> sls = Al.sourcelists
awe> dates = [sl.frame.observing_block.start for sl in sls]
awe> arlist =  ['RA', 'DEC', 'MAG_ISO', 'MAG_AUTO', 'MAG_APER']
awe> r = Al.get_data_on_associates(arlist, count=len(dates))
awe> #for aid in r.keys():      # plots eveything
awe> for aid in r.keys()[:100]: # plots only first 100 stars
....     # index 5 = MAG_AUTO
....     mags = [r[aid][i][5] for i in range(len(r[aid]))]
....     datesmags = zip(dates,mags) # sort by obsdate
....     datesmags.sort()
....     date = [datemag[0] for datemag in datesmags]
....     mag = [datemag[1] for datemag in datesmags]
....     l = pylab.plot(date, mag ,'k.', date, mag, '-')
....
awe> dt1 = datetime.datetime(2005,3,1)
awe> dt2 = datetime.datetime(2005,6,15)
awe> pylab.xlim(dt1, dt2)
\end{verbatim}

\begin{figure}
\begin{center}
\includegraphics[angle=0,width=90mm]{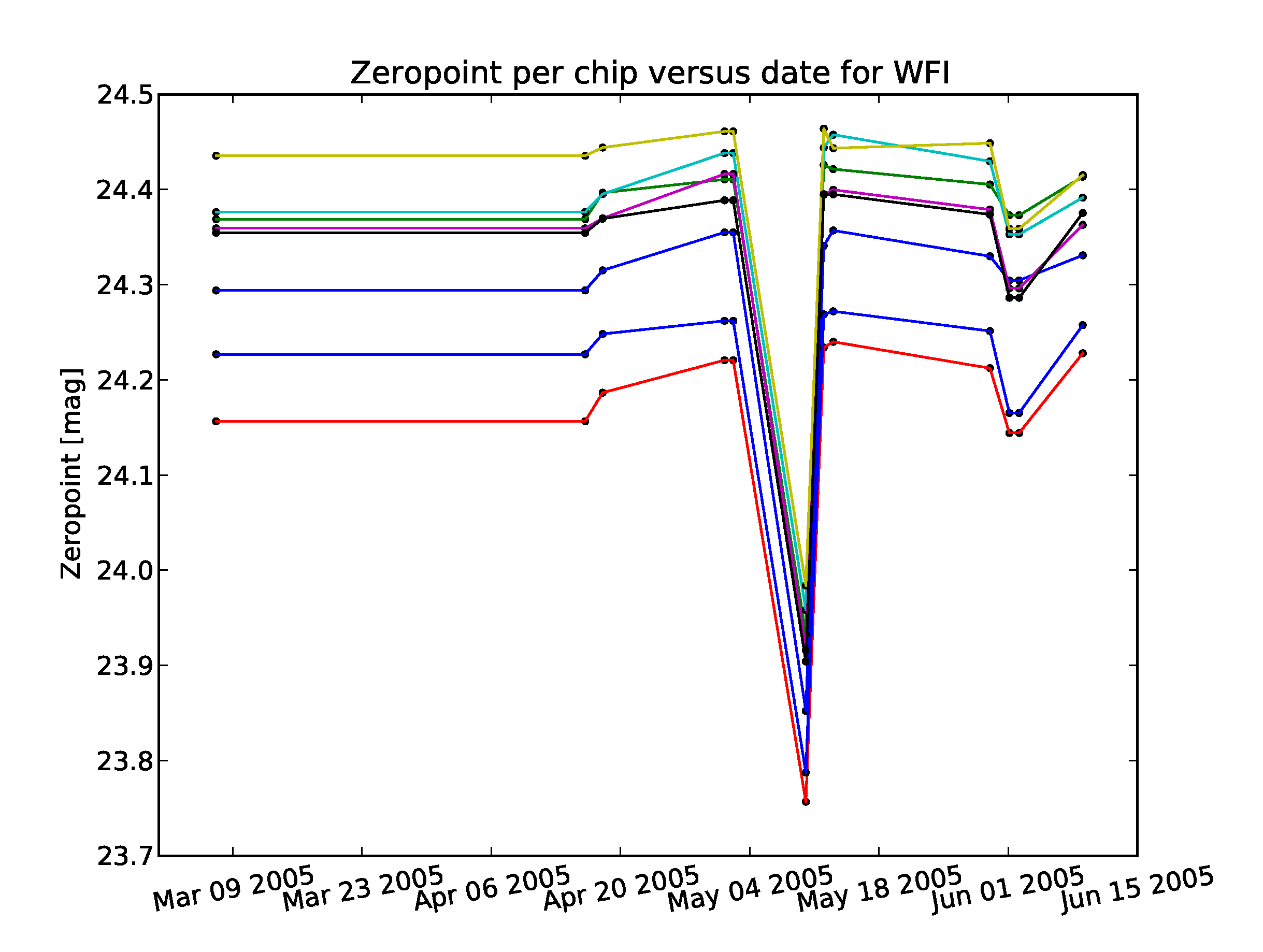}
\caption{
A plot of zeropoint versus the date for all 8 WFI detectors.  The systematic
offsets in zeropoint from night to night is clearly seen.
}\label{fig:zerop}
\end{center}
\end{figure}

In this last example, the zeropoint of each chip is compared
over time with the zeropoints of all the other chips.  The results can be seen in
Fig.~\ref{fig:zerop}.
\begin{verbatim}
awe> for chip in context.get_chips_for_instrument('WFI'):
....     zeropnts = []
....     for sl in sls:
....         for reg in sl.frame.regridded_frames:
....             if reg.chip.name == chip:
....                 red = reg.reduced
....                 break
....         pht = PhotometricParameters.select_for_reduced(red)
....         zeropnts.append(pht.zeropnt.value)
....     dateszps = zip(dates, zeropnts)
....     dateszps.sort()
....     date = [datezp[0] for datezp in dateszps]
....     zeropnt = [datezp[1] for datezp in dateszps]
....     pylab.plot(date, zeropnt, 'k.', date, zeropnt, '-')
....
awe> pylab.xlim(dt1, dt2)
\end{verbatim}
Zeropoint residuals with respect to that of any chip or to the mean zeropoint
per day can easily be obtained with only slight additions to the example code
presented above.  This can give a clearer view of how the zeropoint of the set of chips
evolves over time.

\section{Quality-WISE web service}\label{sec:qwise}

All objects stored in the \aw\ database are stored with their processing and quality
parameters.  These parameters can be accessed in many ways: from the
command-line interface queries, from direct access to the database, or from web
services such as CalTS (\url{calts.astro-wise.org}) or DBView
(\url{dbview.astro-wise.org}).  In \awenv, we have implemented a quality web
service that combines all three methods and collects the most
relevant metadata for the purpose of quality control:
\url{quality.astro-wise.org}.

\begin{figure}
\begin{center}
\includegraphics[angle=0,width=118mm]{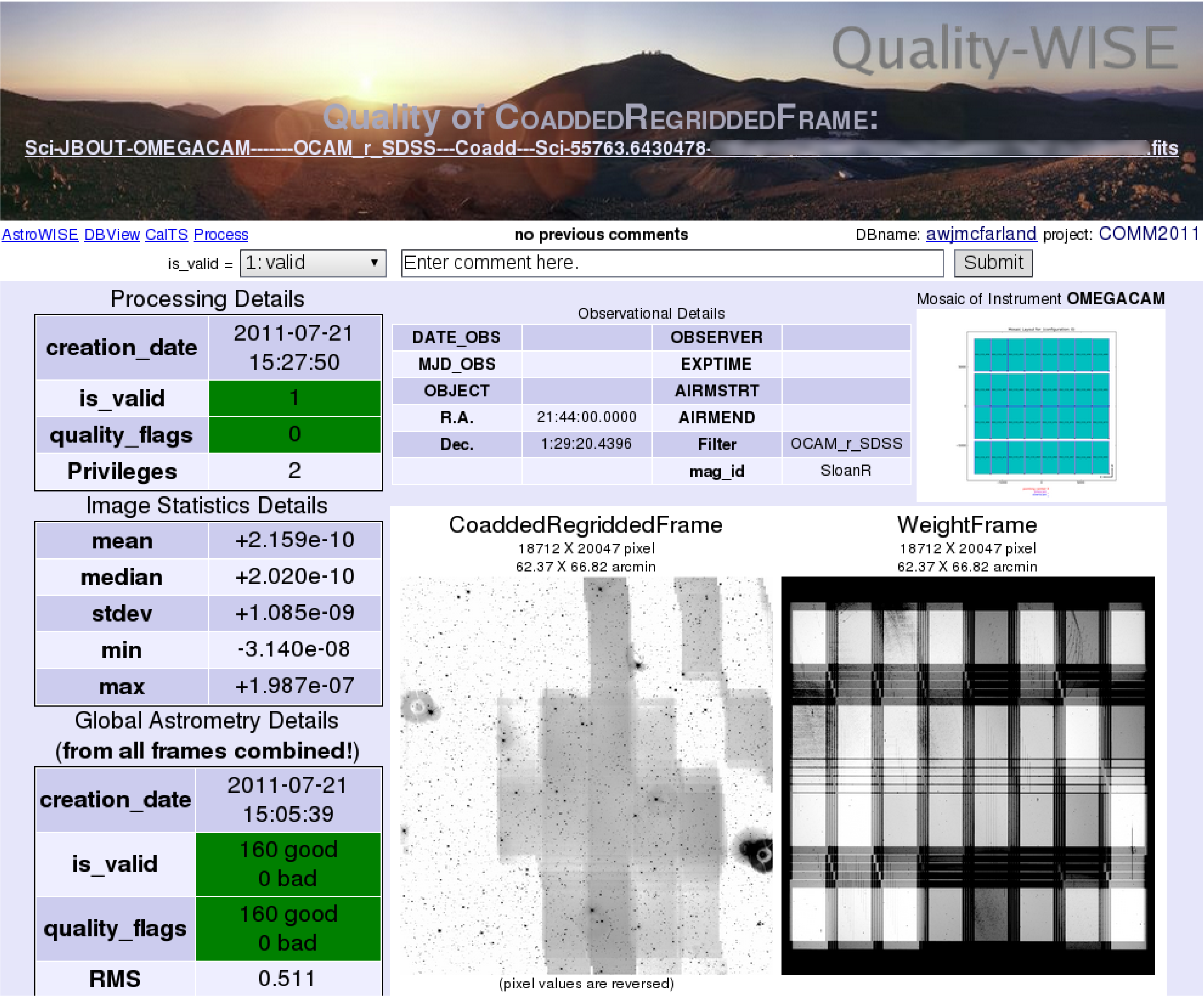}
\caption{
Screen-shot of the upper part of a \qwise\ page.  This view shows the quality of
an OmegaCAM coadded frame.  At the very top is the type of object and a link to
the file on the dataserver (a unique hash value in the filename link is
purposely obscured for security reasons).  Directly below the banner is the top
bar with links and basic actions.  Below this is tabular information about the
object and graphical inspection plots (a thumbnail of the image on the left and
its weights on the right, cf. Fig.~\ref{fig:ithum}).  Note that green fields
indicate values within specified ranges that will be red when out of specified
ranges.
}\label{fig:qwist}
\end{center}
\end{figure}


The \qwise\ interface is accessed primarily through the DBView service by
clicking on the \textsf{\underline{quality}} links associated with science data
objects.  The linked quality pages summarize observational and statistical
details and add a schematic representation of the detector, thumbnails of pixel
data, and various derived inspection plots (see Sect.~\ref{sec:inspe}).
A basic interface is also included to flag or to publish data directly.
Links to the quality pages of associated objects (e.g., progenitor
or derived data products) also exist.  Details of how the \qwise\ service can
be applied to real-world applications can be found in \citet{verdoes}.

\subsection{\qwise\ top bar}

At the top of every \qwise\ page is the class name of the object and a link to
the associated data file on a data server (see Fig.~\ref{fig:qwist}).  There is
a bar below the banner image with links on the left to the \aw\ homepage and to
the database viewer, calibration timestamps and target processor web services.
On the right is the currently logged-in user and project name.  These link to
interfaces to change the user and/or the project via browser cookies.  In the
center, there is an indication of comments associated with the object and an
interface to add comments.  This is typically done when the validity of the
object is changed using the \texttt{is\_valid} interface.  This interface
allows one of 3 levels of validity to be assigned: $0
= invalid$, $1 = valid$ or $2 = publishable$ (see Sect.~\ref{sec:visib}).  Pressing the
\textsf{\underline{Submit}} button stores the validity value and comment, where
applicable, prior to reloading the quality page.  For special purposes such as
surveys, the validity choices can be expanded and the comment interface can have
pre-specified strings included for efficiency.

\subsection{Observational details}\label{sec:obser}

The observational details for the object being inspected are directly below the top bar of a \qwise\ page (see Fig.~\ref{fig:qwist}).  The values are taken
directly from the object stored in the database and include: date of the
observation in human readable and modified Julian date (\texttt{DATE\_OBS} and
\texttt{MJD\_OBS}, respectively), the name of the object observed
(\texttt{OBJECT}), right ascension and declination coordinates (R.A. and Dec.,
respectively), the observer responsible for the observation
(\texttt{OBSERVER}), the exposure time (\texttt{EXPTIME}), the airmass at the
start and end of the observation (\texttt{AIRMSTRT} and \texttt{AIRMEND},
respectively), the filter used for the observation (Filter), and the magnitude
identifier of the filter, i.e., the photometric system (\texttt{mag\_id}).

To the right of the observational details table is a graphical representation
of the detector-plane layout for the individual detectors.  The detectors
highlighted in light blue are those that participated in the current data
object.  In the example of a \coadd\ here, all detectors are highlighted as all
detectors are represented in the data.

\subsection{Processing and statistical details}

On the left side of every \qwise\ page are processing details and
statistics of the main and associated objects (see Fig.~\ref{fig:qwist}).
The main characteristic of this side bar is the highlighting of important
quality parameters (see Table~\ref{tab:proce}).  When a parameter
is within a specified range indicating good quality, the entire cell is colored
green, when the parameter is outside this range, the entire cell is colored
red.  In addition, when the cursor is positioned over any of these cells, the reason for the indicated quality is displayed.

\begin{figure}
\begin{center}
\includegraphics[angle=0,width=118mm]{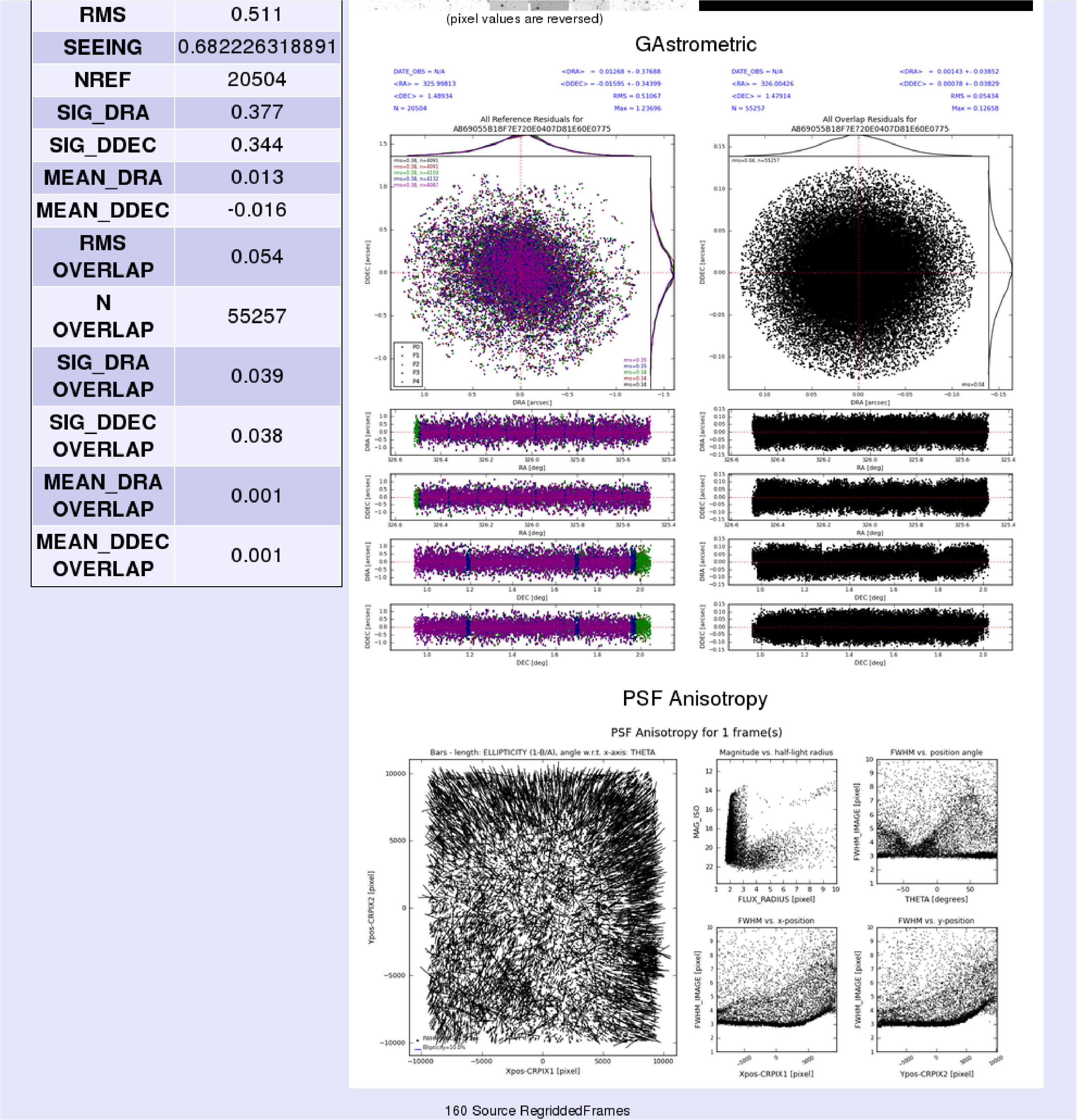}
\caption{
Screen-shot of the middle part of the \qwise\ page shown in Fig.~\ref{fig:qwist}.  
The remainder of the statistical information of the combined global astrometric 
solution can be seen on the left.  The astrometric residuals plots representing 
the quality of the solutions used to make the coadded frame are on the top-right.  
The PSF anisotropy plot for the coadded frame is at the bottom.
}\label{fig:qwism}
\end{center}
\end{figure}

Processing details show when the object was created (\texttt{creation\_date}),
its validity (\texttt{is\_valid}), if any quality flags have been set
(\texttt{quality\_flags}), and to what level it has been published
(Privileges).  See Sect.~\ref{sec:visib} for more on these last three parameters.
Furthermore, statistics of the main object and associated astrometric and photometric objects, if any, are also listed (see also Fig.~\ref{fig:qwism}).

\subsection{Inspection plots}

The main body of each \qwise\ page is dominated by the inspection plots.  These
plots are of the sort described in Sect.~\ref{sec:inspe}.  They always start
with an image thumbnail (with reverse pixel values) and a weight thumbnail
(when applicable) showing lower weights as darker values (see
Fig.~\ref{fig:qwist}).  Below this is the astrometric reference residuals plot
of the individual \textit{reduced frame} local solution, or the astrometric reference and
overlap residuals plots of the composite global solution for \textit{coadded frames} (see Fig.~\ref{fig:qwism}).
In this latter case, the additional plot shows the internal accuracy of the
global solution.  Below the astrometric plots can be the photometric plots showing
the data used to derive the zero point and the results of the illumination
correction derivation (see Figures~\ref{fig:iphot}~and~\ref{fig:iillu}).  These are only shown for non-coadded objects.  The last plot shown is the PSF anisotropy of the
sources in the observation shown at the bottom of
Fig.~\ref{fig:qwism}.

\begin{figure}
\begin{center}
\includegraphics[angle=0,width=118mm]{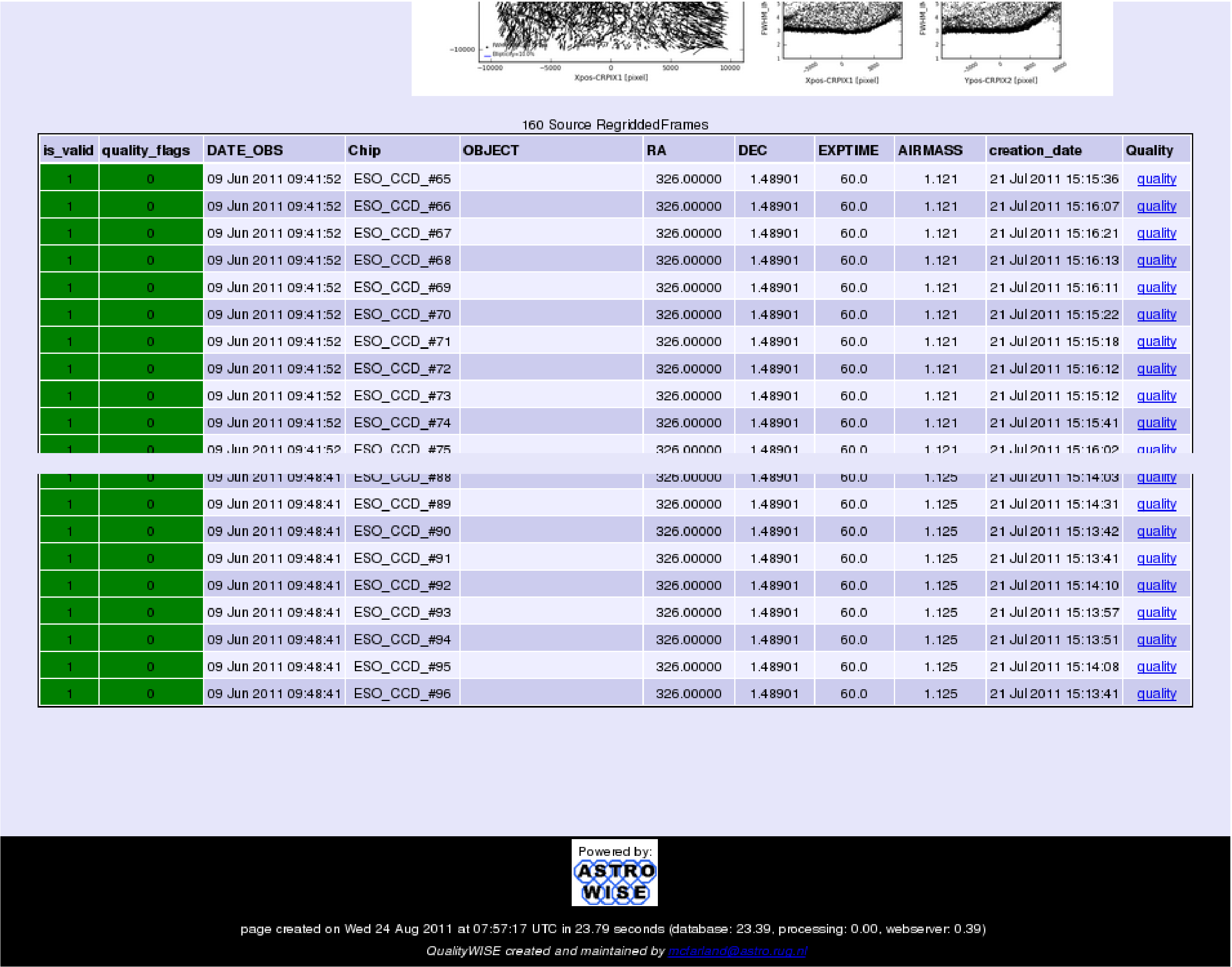}
\caption{
Screen-shot of the lower part of the \qwise\ page shown in Figures~\ref{fig:qwist}~and~\ref{fig:qwism}.
Near the top is the list of progenitor frames'
information.  This list contains 160 entries and is truncated here.  Basic information about the progenitor frames is provided in this
list along with links
to their quality pages (far right).  Page
creation information is presented at the very bottom including a breakdown of creation times into three bins:
database time, processing time and web server time.
}\label{fig:qwisb}
\end{center}
\end{figure}

\subsection{Progenitor/derived quality}

For science data, each data product has progenitor data and derived data.  The
quality pages for these data are linked near the bottom.  In the case of the
\coadd\ quality page in Fig.~\ref{fig:qwisb}, there is only progenitor data.
This consists of a list of 160 {\regri}s.  The information listed is nearly
identical to that described in the observational details table (see
Sect.~\ref{sec:obser}).  At the far right of each entry is the link to the
quality page of the progenitor object.

\section{Summary}\label{sec:summa}

The approach for quality control of astronomical data in the \awinfsys\ has been described.  The comparison to quality control techniques used in other systems has been presented.  It was shown that
the \aw\ approach has advantages for any individual user or group of users in
that it allows the quality to be assessed for not only the final data product,
but also any other progenitor data product in a simple and transparent way
through database linking of all data objects ({\proce}s).

This quality control is built into all aspects of the \aw\ information system.
From the point where raw data enters the system, through all processing steps
to the final data product, quality control mechanisms permeate throughout.
Moreover, the quality of any stage of
data processing can be assessed with quality parameters and inspection plots.

Using metadata (quality- or non-quality-related) stored in all linked objects,
diagnostic plots can be created quickly using a relatively small
amount of command-line code.  This has been shown with examples using archive data from the WFI instrument at La Silla Observatory and (pre-)survey data from the newly commissioned
OmegaCAM instrument at the Paranal Observatory.  The code can be added to
simple scripts for the benefit of the individual user, or eventually find its
way into the core of the system benefiting all users alike.

All the quality control aspects of the \awenv\ have been gathered into a
webservice called \qwise.  This service allows quick viewing of the metadata
and inspection plots of the data in question and of any progenitor or derived
data.  It also provides a simple interface for a user or group of users to
validate data and comment on its quality.

Taken as a whole, the \aw\ approach to quality control is a comprehensive and
efficient method to perform quality checks on individual users' data or on the
data from large astronomical surveys.  It is constantly being updated as newer,
better quality control methods are discovered or derived, and will always stay
on the cutting edge to maintain its advantages.

\begin{acknowledgements}
\aw\ is an on-going project which started from a FP5 RTD programme funded
by the EC Action ``Enhancing Access to Research Infrastructures''. This work is
supported by FP7 specific programme ``Capacities - Optimising the use and
development of research infrastructures''.  Special thanks to Philippe
H\'{e}raudeau and Ivona Kostadinova for their constructive comments.
\end{acknowledgements}


\end{document}